\documentclass[11pt]{article}
\def\be{\begin{equation}}
\def\ee{\end{equation}}
\def\ba{\begin{eqnarray}}
\def\ea{\end{eqnarray}}
\def\la{\langle}
\def\ra{\rangle}
\def\a{\alpha}
\def\b{\beta}
\def\m{\mu}

\def\h{\hskip 1cm}

\def\lo{\longrightarrow}
\def\A1{A_{-1}}
\usepackage{graphicx}
\begin{document}
\begin{titlepage}
\vspace{4cm}
\begin{center}{\Large Transfer of d-Level quantum states through spin chains by random swapping}\\
\vspace{2cm}\h A. Bayat
\footnote{email:bayat@physics.sharif.ac.ir},\h V. Karimipour
\footnote{Corresponding author, email:vahid@sharif.edu}\h  \\
\vspace{1cm} Department of Physics,\\ Sharif University of Technology,\\
P.O. Box 11365-9161,\\ Tehran, Iran
\end{center}
\vskip 3cm
\begin{abstract}
We generalize an already proposed  protocol for quantum state
transfer to spin chains of arbitrary spin. An arbitrary unknown $d-$
level state is transferred through a chain with rather good fidelity
by the natural dynamics of the chain. We compare the performance of
this protocol for various values of $d$. A by-product of our study
is a much simpler method for picking up the state at the destination
as compared with the one proposed previously. We also discuss
entanglement distribution through such chains and show that the
quality of entanglement transition increases with the number of
levels $d$.
\end{abstract}
\end{titlepage}

\section{Introduction}\label{sec1}
Since the proposal of S. Bose \cite{bose} for transferring quantum
states via natural evolution of quantum spin one-half chains, there
have been many types of extensions of this idea in various
directions. For example it has been shown that perfect transfer is
possible for a special class of Hamiltonians, called mirror-periodic
\cite{christandle,Albanese}.  It has also been shown that one can
achieve better fidelities either by using multiple chains
\cite{bose2} or by allowing the parties to have access to more than
one site of the chain \cite{burgarth} or by using chains with longer
range interaction than nearest neighbor \cite{Avellino}. The effect
of thermal fluctuations \cite{bayat} and decoherence
\cite{burgarth3,jian} have also been taken into account. Some other
aspects of this protocol have been studied in
\cite{osborne,subrahamanyan, chiara}. \\
However to our knowledge there has been no attempt to generalize
this proposal to chains of particles of arbitrary spin. The aim of
this paper is to extend this proposal in this new and fundamental
direction. There are good reasons why such an extension is
worthwhile. First, until a particular experimental proposal for
qubit quantum computer is widely accepted as the platform for
implementation of quantum computers, we have to formulate various
theoretical protocols for particles with arbitrary number of levels,
the so-called qudits. In fact for this reason, various protocols of
quantum computation and information, like cloning \cite{clone,Keyl},
cryptography \cite{cryp} and teleportation \cite{tele,tele2} have
been generalized to $d$- dimensional systems. Second, from purely
theoretical point of view we will learn very much in developing a
particular scheme like quantum state transfer in a way such that the
role of dimensionality can be studied in detail. In fact the work of
Bose \cite{bose} can be rephrased in a way which demands such an
extension in a quite natural way: It is well known that a quantum
state can be transferred perfectly through a chain by sequential
application of the swap operator defined as $P|\a,\b\ra=|\b,\a\ra$.
However this method requires control on every qubit throughout the
chain. Instead in \cite{bose} a state is coupled to left hand side
of a spin one-half chain, governed by a ferromagnetic Heisenberg
Hamiltonian $$
    H = - J\sum_{i}{\bf S}_i\cdot{\bf S}_{i+1}+B\sum_{i=1}^N
    S_{iz},\nonumber
$$ where $S_i=\frac{1}{2}\sigma_i$, $\sigma_i$'s are
the Pauli operators, $J>0$ is the coupling constant and $B$ is the
magnetic field. Then the natural evolution of this chain will
transfer the state to the right hand side,  with a good fidelity
provided that the state is extracted at an optimal time. This method
can be named random swapping of a state. The reason is that using
the identity
$$P=\frac{1}{2}(I+\vec{\sigma}\cdot \vec{\sigma}),$$ where $P$ is the permutation operator ($P|\a,\b\ra=|\b,\a\ra$), and a suitable
redefinition of constants, $H$ can be rewritten as
\begin{equation}\label{Heis2}
    H = - J\sum_{i} P_{i,i+1}+B\sum_{i=1}^N S_{iz},
\end{equation}
On the sector with fixed total spin $S_z$, the evolution operator is
equivalent to $U=e^{iJt\sum_{i} P_{i,i+1}}$, where we have set
$\hbar=1$ and ignored an overall phase. Thus for an infinitesimal
time step $\epsilon$, we have
$$|\psi(t+\epsilon)\ra=|\psi(t)\ra+\sum_i iJ\epsilon
P_{i,i+1}|\psi(t)\ra,$$ which shows that the state
$|\psi(t+\epsilon)\ra$ is obtained by adding to $|\psi(t)\ra$ an
equal superposition of states in which the spins of two adjacent
sites have been swapped, hence the name random swapping.
 Thus the result
of \cite{bose} can be rephrased in the following form: for qubits,
random swapping achieves a fidelity which is reasonably good
compared to that of sequential swapping (for which $U=\prod_i
P_{i,i+1}$). In particular when the length of the chain is 4, the
results of \cite{bose} imply that sequential and random swapping
attain almost equal fidelity. Once interpreted in this way, we can
ask naturally what form this
comparison takes for states of arbitrary dimensions. \\
We should note that the Hamiltonian (\ref{Heis2}) can always be
expressed in terms of nearest-neighbor scalar spin interaction terms
although in each dimension it takes a specific form, for example in
dimension $d=3$, it takes the form
$$
H = - J\sum_{i}\left({\bf S}_i\cdot{\bf S}_{i+1}+ ({\bf
S}_i\cdot{\bf S}_{i+1})^2\right)+B\sum_{i=1}^N S_{iz}.
$$
We will find that for a fixed distance, the fidelity decreases with
dimension $d$, but reaches a saturated value depending on the
distance and  when the sender and the receiver are 4 sites apart,
nearly perfect transfer is possible for any dimension $d$. As a
by-product of our study, we will propose a much simpler method for
state transfer, one in which the magnetic field is kept to a
vanishingly small value, instead of tuning it to a
distance-dependent value as in the original protocol of \cite{bose}.
The structure of this paper is as follows. In section \ref{sec2} we
introduce the basic protocol in $d$ dimensions, and derive the basic
relations that we need in the sequel. In section \ref{sec3} we study
the problem of entanglement distribution in such chains.
In section \ref{sec4} we conclude with a discussion. \\

\section{Quantum state transfer in chains of qudits}\label{sec2}
Originally the problem of state transferring was considered for an
open chain\cite{bose}. However in that same work it was shown that
in a ring of size $2N$ one can as efficiently transfer states as in
an open chain as long as the distance between the sender and the
receiver is not longer than $N$. To use the the advantage of
simplicity of eigenfunctions of the Hamiltonian, we consider a
periodic chain of length $N$, where each site comprises a state of a
$d$ level system with basis states $|\mu\ra, \ \ \mu=0,1,\cdots
,d-1$. The evolution of the chain is governed by the Hamiltonian,
\begin{equation}\label{Hamiltonian}
    H=\frac{-J}{2}\sum_{i=1}^N (P_{i,i+1}-1)+B\sum_{i=1}^N S_{iz},
\end{equation}
where the operator $P_{i,i+1}$ is the permutation operator on sites
$i$ and $i+1$, and $S_{iz}$ is a diagonal operator acting on the
states of site $i$ as,
 $S_z|\mu\ra=\mu|\mu\ra,$ for $\mu=0,1,...,d-1.$ Note that $S_z$,
when shifted suitably, plays the role of the third component of the
spin operator. Thus $B$ plays the role of a magnetic field in the
$z$direction.  The Hamiltonian (\ref{Hamiltonian}) reduces to the
Heisenberg Hamiltonian for spin $1/2$ states, and to the
bilinear-biquadratic hamiltonian for spin $1$. For other spins it
contains high power of the term $({\bf S}_i\cdot{\bf S}_{i+1})$. We
assume that $B$ is positive.\\

The ground state of this Hamiltonian is given by
 $|{\bf
0}\ra=|0\ra^{\otimes N}$ with energy  $E_g=0$. \\
The reason is the following. Since the permutation operator has the
property $P^2=I$, its eigenvalues are $\pm 1$, and the operator
$J(1-P_{i,i+1})$ will be a positive operator with eigenvalues $0$
and $2J$. Therefore in the absence of magnetic field, the
Hamiltonian, being a sum of positive operators, is positive and
since the states $|\mu\ra^{\otimes}$, $\mu=0,\cdots d-1$ all have
zero energy, they form the degenerate ground state of $H(B=0)$. The
magnetic field only removes the degeneracy and lowers the energy of
the state $|0\ra^{\otimes N}$, with respect to others (note that in
our notation $|0\ra$ has the lowest value of spin component.) Since
the Hamiltonian commutes with $S_z$, and $H(B=0)$ can be
diagonlaized in sectors with fixed $z$ component of spin, this
argument is
valid for all values of the magnetic field $B$.\\

We should stress that in the absence of magnetic field, the phase
diagram (i.e. the character and long range order in the ground
state) of (\ref{Hamiltonian}), may be quite complicated. This will
then affects crucially the quality of state transfer in such chains,
a problem which has been recently investigated for spin 1 chains in
\cite{kai}. In the presence of magnetic field however, the ground
state has a simple ferromagnetic order given by the ground state
$|0\ra^{\otimes N}$.\\

Let us denote a state in which the $i$-th site has been exited to
the level $\mu$ by $|\mu_i\ra$, i.e. $$|\mu_i\ra = |0,\cdots 0,
\mu,0,\cdots 0\ra.$$ The permutation operators in $H$ only displace
this state through the chain and hence the Hamiltonian can be
diagonalized in each sector in which the number and type of excited
local states is fixed. This is a consequence of a number of
conservation laws, namely
$$[H,Q^{m}]=0 \ \ \ \ \ ,\ \ \ \ \  Q^{(m)}:=\sum_{i=1}^N(S_{z,i})^m$$ for $m=1,2,\cdots ,d-1.$ In $d=2$
dimensions only the $Q^{(1)}$ charge is conserved. These
conservation laws imply for example that a state like, $|1,1,\cdots
0, 0, 0\ra $ can not evolve to a state like $|2,0,0\cdots,
0,0,0\ra$, since although their $Q^{(1)}$ charge are
equal they have different $Q^{(2)}$ charges. \\
The states with only one site excited are called one particle
states and the subspace spanned by these vectors comprise the
one-particle sector of the full Hilbert space. Let us denote by
$V_1^{(\mu)}$ the one particle sector with $Q^{(1)}$ charge equal
to $\mu$. The whole one particle sector is $$V_1=V_1^{(1)}\oplus
V_1^{(2)}\oplus \cdots V_1^{(d-1)}.$$ The Hamiltonian can be
diagonalized in $V_1^{(\m)}$ with eigenvectors given by,
$$
    |E^m_\mu\ra=\frac{1}{\sqrt{N}}\sum_{k=1}^Ne^{\frac{i2\pi
    km}{N}}|\mu_k\ra, \h m=1,2,...,N,\nonumber
$$
with energy  given by $E^\mu_m=J-Jcos(\frac{2\pi
    m}{N})+B\mu.$
For quantum state transferring we can consider site $s$ as the
sender of the system and site $r$ as a receiver. The initial state
that should be sent is
$$|\psi_s\ra=\sum_{\mu=0}^{d-1}a_\mu|\mu\ra.$$ So the initial state
of the system (the site $s$ plus the chain) is,
$$
    |\psi(0)\ra=|\psi_s\ra\otimes|\textbf{0}\ra=a_0|\textbf{0}\ra + \sum_{\mu=1}^{d-1}
    a_{\mu}|\mu_s\ra.\nonumber
$$
In view of the fact that $H|\textbf{0}\ra = 0 $, the state at time
$t$ will be,
$$
    |\psi(t)\ra=a_0|\textbf{0}\ra + \sum_{\mu=1}^{d-1}
    a_{\mu}e^{-iHt}|\mu_s\ra=a_0|\textbf{0}\ra + \sum_{k=1}^N\sum_{\mu=1}^{d-1}f_{ks}a_{\mu} e^{-iB\mu
    t}|\mu_k\ra.
$$
In deriving this formula we have used the fact that $[S_z,
\tilde{H}:=\sum_{i}P_{i,i+1}]=0$ and the conservation laws which
restricts the evolution to the one particle sector of fixed
$Q^{(1)}$ charges. We have also defined
$$
    f_{ks}:=\la\mu_k|e^{-i\tilde{H}t}|\mu_s\ra,
$$
which is indeed independent of $\mu$ and hence can be taken outside
the sum.

The state of site $r$ which is acting as the receiver will be
generally mixed, so is denoted by $\rho_r(t)$ and is obtained by
tracing out the other sites.
\begin{eqnarray}
    \rho_r(t)&=&tr_{\hat{r}}{|\psi(t)\ra \la
    \psi(t)|}=(1-\sum_{\mu=1}^{d-1}|a_{\mu}|^2|f_{rs}^{\mu}|^2)|0\ra\la 0|+\sum_{\mu=1}^{d-1}a_0a_\mu^*f_{rs}^{\mu*}|0\ra\la\mu|\cr
    &+&\sum_{\mu=1}^{d-1}a_0^*a_\mu f_{rs}^{\mu}|\mu\ra\la0|+\sum_{\mu,\nu=1}^{d-1}a_\mu a_\nu^* f_{rs}^{\mu}f_{rs}^{\nu *}
    |\mu \ra \la \nu|. \nonumber
\end{eqnarray}
Rearrangement of the right hand side yields
$$
    \rho_r(t)=(1-P)|0\ra \la 0| + P |\phi\ra \la \phi|,
$$
where
$$
 P=|f_{rs}|^2(1-|a_0|^2)+|a_0|^2,
$$
and
$$
    |\phi\ra =\frac{1}{\sqrt{P}}( a_0|0\ra + f_{rs}\sum_{\mu=1}^{d-1}a_{\m} e^{-iB\mu
    t}|\mu\ra). \nonumber
$$
Alternatively we can say that the input state $\rho_s(0)$ is mapped
to the output state $\rho_r(t)$ by the positive map $$\rho_r(t) =
\sum_{\mu}A_{\mu}\rho_s(0) A_{\mu}^{\dagger},$$ where the so-called
Kraus operators $A_{\mu}$ are given by
\begin{equation}\label{Krausdec}
    A_0=|0\ra\la 0|+\sum_{\mu=1}^{d-1}f_{rs}^\mu \ \
    |\mu\ra\la\mu |,\ \ \ \
    A_\mu =\sqrt{1-|f_{rs}^\mu |^2} \ \ \ |0\ra\la\mu|, \ \ \ \
    \ \mu=1,...,d-1.
\end{equation}
The fidelity between the received state $\rho_r(t)$ and the initial
state $\rho_s(0)=|\psi_s\ra\la \psi_s|$ is defined by
$F=|\la\psi_s|\rho_r(t)|\psi_s\ra|^2$ which turns out to be
$$
    F(t)=|a_0|^2+\sum_{\mu=1}^{d-1}|a_0|^2|a_\mu|^2\{1-|1-f_{rs}^\mu|^2
    \}+\sum_{\mu,\nu=1}^{d-1}|a_\mu|^2|a_\nu|^2f_{rs}^\mu
    f_{rs}^{\nu*}.\nonumber
$$
In the sequel we should maximize this fidelity when it is uniformly
averaged over the input states. The average is defined by
$$
\la F(t)\ra=\int F(t) dU,\nonumber
$$
where $dU$ is an invariant (Haar) measure over the $SU(d)$ group,
normalized such that $\int dU =1$. The reason for this choice of
measure is as follows. Let us fix a basis, like $\{|0\ra,
|1\ra,\cdots |d-1\ra\}$. We take a fixed reference state like
$|0\ra$ and note that every arbitrary state $|\psi\ra$ can be
obtained from $|0\ra$ by the action of a unitary operator $U$, i.e.
$|\psi\ra=U|0\ra$, for some non-unique $U\in SU(d)$. However, any
two unitary matrices $U$ and $Ug$, where $g\in SU(d-1)$ leaves
$|0\ra$ invariant, lead to the same state $|\psi\ra$. Therefore a
proper measure that prevents this multiple counting is a measure
over $U(d)/U(d-1)$. However since every state is multiply counted
equally ( by a factor which is exactly the volume of the group
$SU(d-1)$), this does not affect the final averaging and we can use
the simple measure over $U(d)$. Invariance of this measure under
left multiplication, i.e. $dU=d(gU)$ guarantees uniformity of the
measure over the space of all states. In two dimensions one can
avoid multiple counting in a simple way, since in this case
$U(2)/U(1)\sim SO(3)/SO(2)\sim S_2$ and therefore, one can use the
measure over the 2 dimensional (Bloch) sphere to count every state
once. This is the
measure used by Bose in \cite{bose}.\\
For a $d-$ dimensional normalized state $|\psi\ra =
\sum_{\mu=0}^{d-1}a_{\mu}|\mu\ra$ an invariant measure yields
trivially $\la |a_{\mu}|^2\ra = \frac{1}{d}\ \ \forall \mu$. To
calculate the other averages we use $|\psi\ra = U|0\ra$ and write
$$
   \la |a_\mu|^4\ra = \la |a_0|^4\ra = \int |U_{00}|^4 dU =
   \frac{2}{d(d+1)},\ \ \ \forall \ \ \mu,
$$
where for the last equality we have used a result from \cite{aubert}
on invariant integration on unitary groups. We can now calculate
$\la |a_\mu|^2|a_\nu|^2\ra$ for $\mu\ne \nu$. In view of the
normalization of the state, we have \begin{eqnarray}
   1&=&\sum_{\mu,\nu}\la |a_\mu|^2|a_\nu|^2\ra=\sum_{\mu}\la
    |a_\mu|^4\ra + \sum_{\mu\ne\nu}\la |a_\mu|^2|a_\nu|^2\ra\cr
    &=& d\frac{2}{d(d+1)}+d(d-1)\la |a_\mu|^2|a_\nu|^2\ra,\nonumber
\end{eqnarray} thus we find
$$
    \la |a_\mu|^2|a_\nu|^2\ra_{\mu\ne \nu}=\frac{1}{d(d+1)}.\nonumber
$$
Using these results we can now calculate the average of fidelity
over a uniform ensemble of input states.
 \begin{equation}\label{avefid}
    \la F(t)\ra =\frac{1}{d}+\frac{1}{d(d+1)}\sum_{\mu=1}^{d-1} ({f_{rs}^{\mu}+{f^*}_{rs}^{\mu}})+
    \frac{1}{d(d+1)}\sum_{\mu,\nu =
    1}^{d-1}f_{rs}^{\mu}{f^*}_{rs}^{\nu}.
\end{equation}
In order to write $\la F(t)\ra $ in a simple form we note that
$$
    |\sum_{\mu=1}^{d-1} f^\mu_{rs}|=|\sum_{\mu=1}^{d-1}e^{-i\mu
    Bt}f_{rs}|= |f_{rs}|\Gamma_d(Bt),
$$
where
$$ \Gamma_d(Bt):=\frac{\sin \frac{(d-1)Bt}{2}}{\sin
    \frac{Bt}{2}},$$
 and
$$
    \sum_{\mu=1}^{d-1} (f^\mu_{rs}+{f^*}^{\mu}_{rs})= 2\cos
    (\gamma_{rs}-\frac{dBt}{2})\Gamma_d(Bt),\nonumber
$$
where $f_{rs}=|f_{rs}|e^{i\gamma_{rs}}$. Inserting these in
 (\ref{avefid}) we find that
\begin{equation}\label{FinalAverageFidelity}
    \la F(t)\ra = \frac{1}{d}+\frac{2}{d(d+1)}|f_{rs}|\cos (\gamma_{rs}-\frac{dBt}{2}) \Gamma_d(Bt) +
    \frac{1}{d(d+1)}|f_{rs}|^2 \Gamma_d^2(Bt).
\end{equation}
For $d=2$ we recover the formula of \cite{bose}.\\

It remains to calculate the explicit expression for the amplitudes
$f_{rs}^{\mu}$. We note that
$$
    f_{rs}=\la\mu_r|e^{-i\tilde{H}t}|\mu_s\ra=\sum_{k=1}^{N}\la\mu_r|e^{-i\tilde{H}t}|E^{\mu}_k\ra\la
    E^{\mu}_k|\mu_s\ra=\frac{e^{-iJt}}{N}\sum_{k=0}^{N-1} e^{iJtcos(\frac{2\pi k}{N})} e^{\frac{i2\pi
    k(r-s)}{N}}.\nonumber
$$
For large $N$, a closed formula for $f_{rs}$ can be obtained by
writing the right hand side as an integral. Thus
$$
    f_{rs}\approx \frac{e^{-iJt}}{2\pi}\int_0^{2\pi} e^{iJtcos(\theta)+
    i\theta(r-s)}= e^{-i(Jt-\frac{\pi}{2}(r-s))} J_{r-s}(Jt),\ \ \ {\rm {Large}}\ \  N,\nonumber
$$
where $J_n$ is the Bessel function of the first kind of order $n$.
It now remains to follow a definite strategy for picking up the
state at the destination point $r$. As is clear from equation
(\ref{FinalAverageFidelity}), there is a distinctive difference
between dimension $d=2$ and any other dimension, since in $d=2$ we
have $\Gamma_2(Bt)=1$ and the magnetic field enters only in one
single term, namely the argument of cosine function. (Note that we
can always re-scale the other coupling constant $J$ to 1.)  Thus in
$d=2$ there is rather a unique strategy, first suggested in
\cite{bose}: at any given time $t$ one finds the magnitude of
$B=B(t)$ which maximizes the cosine function to unity and then
searches among the values of time $t$ to determine the optimal time
$t_{opt}$ for picking up the state. This will then determine the
optimal value of the magnetic field through the relation
$B_{opt}=B(t_{opt}).$ Note that since $t_{opt}$ depends on the
distance $r-s$ between the sender and the receiver, the optimal
value of the magnetic field also depends on this distance. This is
an inconvenient feature of this strategy.

A by-product of the present work is that a much simpler strategy can
be used, namely: tune the magnetic field to a vanishingly small
value, then the optimal time for picking up the state is almost
independent of the magnetic field.  To see this we note that for
higher values of $d$, the magnetic field enters in two different
ways in the final formula for the average fidelity, namely in the
argument of the cosine function as in $d=2$ and in the function
$\Gamma_d(Bt)$. These two functions may have incompatible properties
so that they may not be maximized simultaneously. The latter
function is maximized when its argument $Bt$ goes to zero, while the
former function has a complicated dependence on $t$ and $B$
separately. Thus one can follow two different strategies for picking
up the states at point $r$, the first one is exactly the same as in
\cite{bose}, explained above. We can also follow a second much
simpler strategy, which has the advantage of no need for
distance-dependent tuning of the magnetic field. We simply apply a
vanishingly small magnetic field ($Bt<<1$) for all the times
involved in the transfer process. On the other hand $B$ should not
be vanishing so that we have a unique ferromagnetic ground state.
 This maximizes the function
$\Gamma_d(Bt)$ to $\approx (d-1)$. We are now left with a function
which is entirely a function of $t$ and can find for any
distance $r-s$, the optimal time and the maximum average fidelity.\\
In figure (1), we show the average fidelity for transferring $d=2,
3,$ and $4-$ level states through half way distances in closed
rings, using this strategy and compare it with the original Bose
strategy. Thus when $r-s=20$,
we are using a ring of size $N=40$. \\

There are a few interesting features. First it is seen that the
average fidelity is almost the same in both strategies, which
implies that we can always, even for $d=2$, use the second method
which is much simpler and does not rely on distance-dependent tuning
of magnetic fields. Second we note that the strong similarity of the
fidelity curves in various dimensions. To see the reason of this
universality, we note that for very small magnetic fields, $Bt<<1$,
where $\Gamma_d(Bt)\sim (d-1)$, the average fidelity behaves as

\begin{equation}\label{FinalAverageFidelity2}
    \la F(t)\ra \sim \frac{1}{d}+\frac{2 (d-1)}{d(d+1)}|f_{rs}|\cos (\gamma_{rs}) +
    \frac{(d-1)^2}{d(d+1)}|f_{rs}|^2.
\end{equation}
The curves in figure (1) show the fidelities at the optimal time
$t_{op}$ where $\la F(t_{opt})\ra$  becomes a maximum, obtained by
numerical searches in a time span $t\in (0,400)$. Let us now suppose
that the optimal time is the time where $\cos
(\gamma_{rs})\approx1$. Since $f_{rs}$ is independent of $d$, we can
obtain a universal relation for optimal fidelities from the above
equation by rewriting it as follows. First we note from the above
equation that
$$ F^{*}_{opt}:=\lim_{d\lo \infty} \la F_{opt}\ra = |f_{rs}|^2.$$
Rearranging the terms of (\ref{FinalAverageFidelity2}) after setting
$\cos \gamma_{rs}\approx 1$, we find
\begin{equation}\label{scaling}
\frac{\sqrt{d(d+1)\la F_{opt}\ra-d}-1}{d-1}=\sqrt{F^{*}_{opt}},
\end{equation}
which is a constant. To check this assumption and the resulting
universality, we draw in figure (2) the left hand side of
(\ref{scaling}) (as obtained from numerical searches for the optimal
time, leading to figure (1) and not by setting $\cos
\gamma_{rs}\approx 1$) for several values of $d$. The
universal behavior is now completely evident. \\
Finally, we note that almost perfect state transfer of any $d-$
level state is possible when $|r-s|=1,2,4$. This possibility of
almost perfect state transfer was first noticed in \cite{bose} for
$d=2$ level states. We now see that this is a general and curious
feature of random swapping for any $d-$ level state. We know that by
sequential swapping at any two consecutive sites, one can perfectly
transfer an unknown state through a chain. However this requires a
multitude of control operations at all sites of the chain. The above
result about perfect transfer over distances of $4$ sites by random
swapping (induced by the natural Hamiltonian dynamics), means that
one can transfer states perfectly over long distances in a chain by
a smaller number of control operations, namely by 1/4 of the number
of sites of the chain. \\
\begin{figure}\label{fig01}
\centering
   \includegraphics[width=14cm,height=10cm,angle=0]{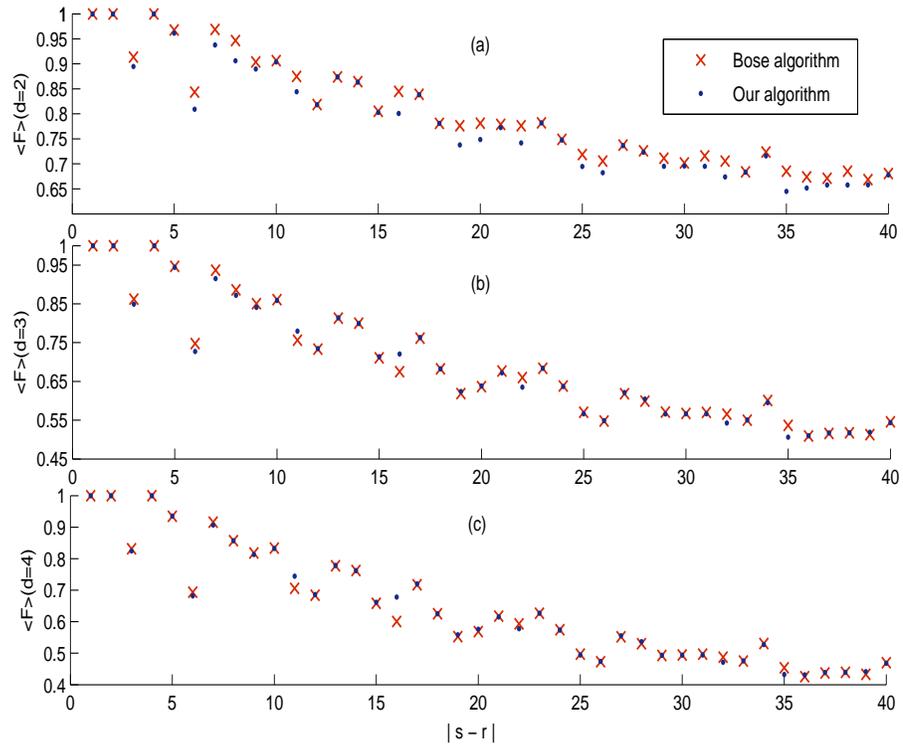}
   \caption{(Color online) The average fidelity for transferring d=$2,3$ and $4-$ level states for two different strategies explained in the text.
   In each case the distance between the sender (s) and the receiver (r) is half the length of the chain
   (N).
   }\label{newFig1}
\end{figure}
Figure (3) shows average fidelities for three different distances,
namely $|r-s|=4,7,14$, as functions of the number of levels $d$. It
is seen that the average fidelity decreases with $d$ and saturates
to a constant value, depending on the distance. Plots $a$ and $b$
refer to two different strategies discussed above.

\begin{figure}
\centering
   \includegraphics[width=14cm,height=10cm,angle=0]{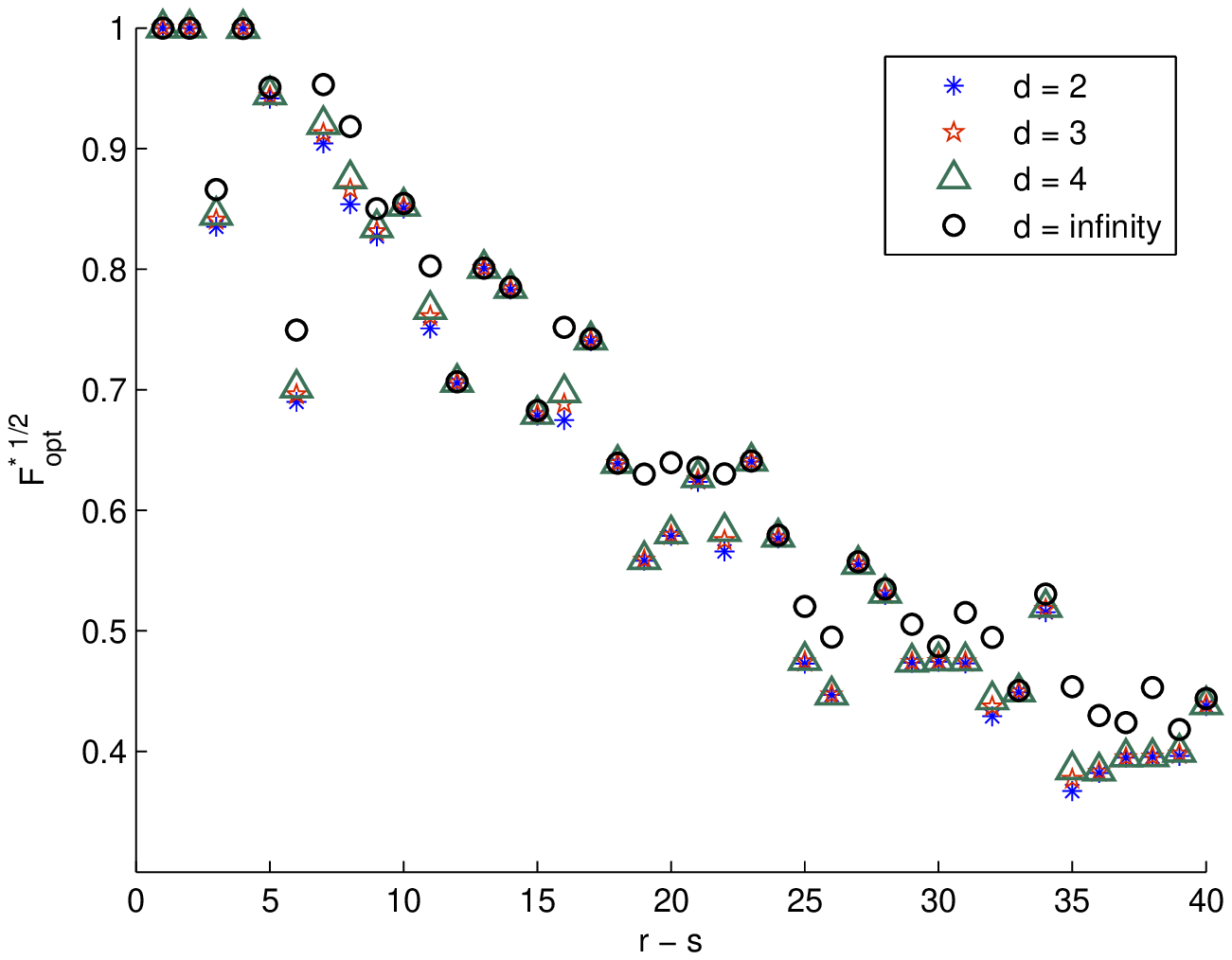}
   \caption{(Color online) The universal relation for optimal fidelities, explained in equation (\ref{scaling}).
   }\label{newFig2}
\end{figure}

\begin{figure}\label{fig2}
\centering
   \includegraphics[width=10cm,height=10cm,angle=0]{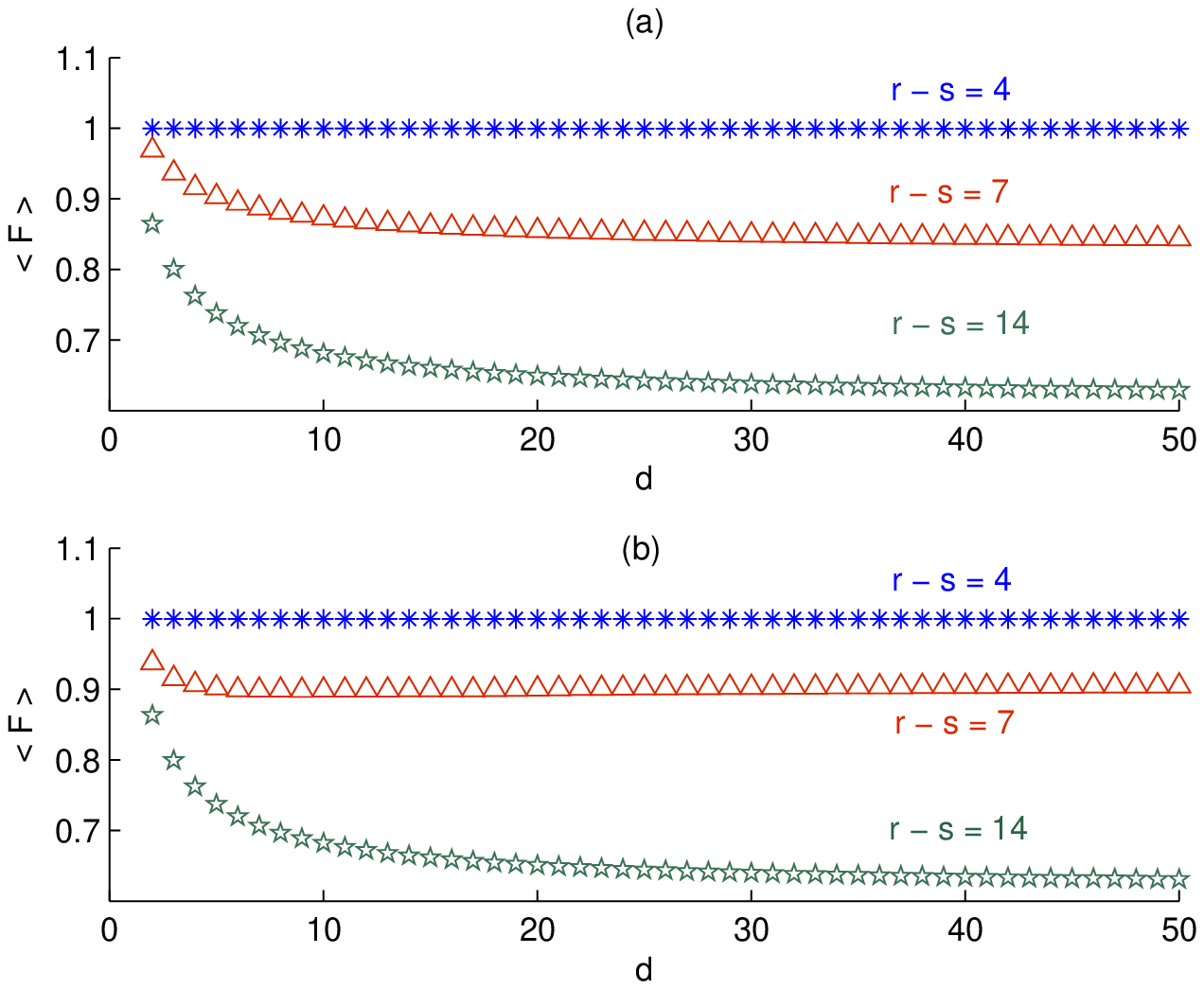}
   \caption{(Color online) The average fidelity for three different
distances, namely $|r-s|=4,7,14$, as functions of the number of
levels $d$. Plots $a$ are for the strategy of Bose \cite{bose} and
plots $b$ are for our strategy.}
    \label{newFig2}
\end{figure}
\section{Entanglement distribution}\label{sec3}
One of the major problems of quantum information science and
technology is the distribution of entangled pairs over long
distances. For flying qubits, such pairs in the form of
polarization-entangled photons have been distributed to various long
distances through optical fibres and free air \cite{gisin, ursin,
jennewein, aspelmeyer, vienna1, vienna2}. For small distances in a
quantum computer, for which we supposedly will be dealing with solid
state devices or ion traps in the future, one needs to distribute
entangled pairs through such chains of qubits. In \cite{bose} a
method was proposed for such a task, when the particles have spin
$1/2$ or have two levels (qubits). Here we generalize this idea to
$d-$ level states or qudits. We will see here that the quality of
entanglement transfer is better for higher
values of $d$.\\

Suppose that a maximally entangled state is prepared between sites
$0$ (not coupled to the chain) and site $s$ of the chain. We want to
use the natural dynamics of the chain to transfer this entanglement
through the chain. In particular we want to see what will be the
entanglement between sites $0$ and $r$ at time $t$. The initial
states is
\begin{equation}\label{maxEn}
    |\psi_{ME}\ra_{0,s}=\frac{1}{\sqrt{d}}\sum_{\mu=0}^{d-1}|\mu \mu
    \ra.
\end{equation}
At time $t$, the joint state of sites $0$ and $r$ can be
determined by the extension of the map (\ref{Krausdec}) to one
acting on the chain and the external site $0$:
\begin{equation}\label{Kraus_En}
    \rho_{0,r}(t)\ra=\sum_{\mu=0}^{d-1} (I\otimes A_\mu) |\psi_{ME}\ra\la
    \psi_{ME}| (I\otimes A_\mu^\dag).
\end{equation}
Insertion of the Kraus operators $A_\mu$s from (\ref{Krausdec}) in
(\ref{Kraus_En}) gives,
\begin{eqnarray}\label{rho_t}
    \rho_{0r}(t)&=& \frac{1}{d} \{ \ \ |00\ra \la 00|+\sum_{\mu=1}^{d-1}f_{rs}^{\mu *}|00\ra \la \mu \mu |
    +\sum_{\mu=1}^{d-1}f_{rs}^{\mu}|\mu \mu\ra \la 00| \cr
    &+&\sum_{\mu,\nu=1}^{d-1}f_{rs}^{\mu}f_{rs}^{\nu *}|\mu \mu\ra \la \nu
    \nu|+\sum_{\mu=1}^{d-1}(1-|f_{rs}^{\mu}|^2)|\mu 0\ra \la \mu 0 | \ \
    \}.\nonumber
\end{eqnarray}
We use Logarithmic Negativity (LN) \cite{vidal} as a measure of
entanglement of the state $\rho_{0r}$ which is defined as,
\begin{eqnarray}\label{LN}
    LN(\rho_{12})=log_2(\|\rho_{12}^{T_2}\|), \ \ \ \ \|O\|&=&tr\sqrt{O^\dag
    O},
\end{eqnarray}
where by the superscript ${T_2}$, the partial trace over the second
space is implied. Logarithmic negativity is an entanglement monotone
which is additive and does not increase on the average under all
partial transpose preserving operations \cite{plenioneg}. For a pure
maximally entangled state like
(\ref{maxEn}), equation (\ref{LN}) yields $LN(|\psi_{ME}\ra)=log_2d$.\\
In order to calculate logarithmic negativity we need eigenvalues of
$\rho_{0r}^{T_r \dag}\rho_{0r}^{T_r}$.  Straightforward calculations
shows that
\begin{eqnarray}\label{rho_t2}
    \rho_{0r}^{T_r \dag}\rho_{0r}^{T_r}&=& \frac{1}{d^2} \{ |00\ra \la 00|+\sum_{\mu=1}^{d-1}|f_{rs}^{\mu}|^2
    |\mu 0\ra \la \mu 0|+\sum_{\mu=1}^{d-1}|f_{rs}^{\mu}|^2
    |0 \mu\ra \la 0 \mu|\cr
    &+&\sum_{\mu,\nu=1}^{d-1}|f_{rs}^{\mu}|^2|f_{rs}^{\nu}|^2|\mu \nu\ra \la \mu
    \nu|+\sum_{\mu=1}^{d-1}(1-|f_{rs}^{\mu}|^2)^2|\mu 0\ra \la \mu 0 |\cr
    &+&\sum_{\mu=1}^{d-1}(1-|f_{rs}^{\mu}|^2)f_{rs}^\mu |\mu 0\ra \la 0 \mu |+
    \sum_{\mu=1}^{d-1}(1-|f_{rs}^{\mu}|^2)f_{rs}^{\mu *} |0 \mu \ra \la \mu 0
    |\}.\nonumber
\end{eqnarray}
It is readily seen that this operator is the direct sum of
two-dimensional  matrices of the form
$$
   \frac{1}{d^2}\left(%
\begin{array}{cc}
  |f_{rs}^\mu|^2 & (1-|f_{rs}^\mu|^2)f_{rs}^{\mu *} \\
  (1-|f_{rs}^\mu|^2)f_{rs}^{\mu} & 1-|f_{rs}^\mu|^2+|f_{rs}^\mu|^4 \\
\end{array}%
\right)\nonumber
$$
in the subspaces spanned by $|\mu 0\ra$ and $|0 \mu\ra$ (with
eigenvalues  $\frac{1}{d^2}, \frac{|f_{rs}^\mu|^2}{d^2}$ ) and a
diagonal matrix spanned by the rest of basis vectors. Putting these
together we find the spectrum of the matrix $\sqrt{\rho_{0r}^{T_r
\dag}\rho_{0r}^{T_r}}$ as follows,
$$
    \left\{%
\begin{array}{ll}
    \frac{1}{d}, & \ \ \ d \\
\frac{|f_{rs}^\mu||f_{rs}^\nu|}{d}, & \ \ \ (d-1)^2 \\
\frac{|f_{rs}^\mu|^2}{d}, & \ \ \ d-1,\\
\end{array}%
\right.
$$
where the number in front of each eigenvalue denotes its degeneracy.
Therefore the logarithmic negativity of the final state between
sites $0$ and $r$ can be computed easily. It is found that
$$
     LN(\rho_{0r}(t))=log_2\{ 1+\frac{1}{d}\sum_{\mu ,\nu =1}^{d-1}|f_{rs}^\mu||f_{rs}^\nu|
    +\frac{1}{d}\sum_{\mu=1}^{d-1}|f_{rs}^\mu|^2\}\nonumber
$$
and because $|f_{rs}^\mu|$ is independent of $\mu$ so we can
simplify the above formula,$LN(\rho_{0r}(t))=\log_2\{
1+|f_{rs}|^2(d-1)\}$. This equation shows that with increasing $d$
the logarithmic negativity and hence the entanglement increases and
indeed approaches its maximum value for continuous variable states.
We can define the efficiency of entanglement distribution as a
measure of the the percentage of entanglement that is gained after
distribution of the maximally entangled state through the chain, so
we introduce the efficiency as, $$
    E=\frac{LN_2}{LN_1}=\frac{\log_2\{
    1+|f_{rs}|^2(d-1)\}}{log_2d}=\log_d\{
    1+|f_{rs}|^2(d-1)\}.\nonumber
$$ Figure (4) shows the efficiency of entanglement of
sites $1$ and $30$ in a ring of size $N=60$ as a function of time
for three different values of $d$.
\begin{figure}\label{fig3}
\centering
   \includegraphics[width=10cm,height=8cm,angle=0]{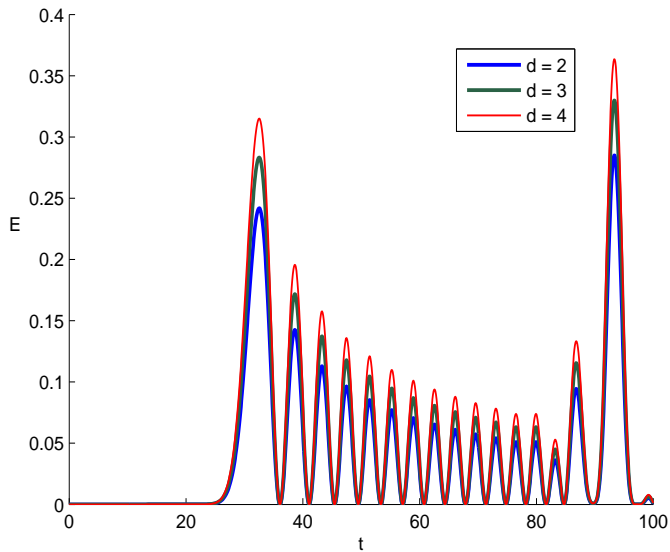}
   \caption{(Color online) Entanglement transition through a chain. The efficiency of transmission of maximally entangled pairs
   through a distance of 30 sites in a chain of length 60 for different $d-$ level
   states. The curves from bottom to top correspond to $d=2$, $d=3$
   and $d=4$.
   }
    \label{newFig2}
\end{figure}
It is seen that the optimal time for picking up the states is
independent of $d$ and the efficiency is increased by increasing
the dimension $d$.
\section{Summary}\label{sec4}
We have generalized the protocol of \cite{bose} for quantum stater
state transfer of qubits to transfer of $d$-level states. On the
theoretical side, we can consider the results of \cite{bose} and the
present paper as an answer to the question "with what fidelity can a
quantum state be transferred through a chain if we use random
swapping instead of sequential swapping?". The latter method is
known to achieve unit fidelity but requires local control at every
site of the chain. We have shown that 1- the fidelity decreases with
the dimension $d$, but reaches a saturated value depending on the
distance, and 2- that when the sender and the receiver are 4 sites
apart, nearly perfect transfer is possible for any dimension $d$. A
by-product of our study is that we have proposed a much simpler
method for state transfer, one in which the magnetic field is kept
to a vanishingly small value, instead of tuning it to a
distance-dependent value as in the original protocol of
\cite{bose}.\\ Furthermore the concept of entanglement distribution
has been studied for $d$-level states and the interesting result is
that the quality of entanglement distribution will be improved by
increasing the dimension $d$.
\section{Acknowledgement}
We would like to thank the members of the quantum information group
in the physics department of Sharif university for very valuable
discussions.


\begin{thebibliography}{}
\bibitem{bose} S. Bose, Phys. Rev. Lett. 91, 207901 (2003).
\bibitem{christandle} M. Christandl, N. Datta, A. Ekert, A.J.
Landahl, Phys. Rev. Lett. 92, 187902 (2004).
\bibitem{Albanese} C. Albanese, M.Christandl, N. Datta, A. Ekert, Phys. Rev. Lett. 93,230502 (2004).
\bibitem{bose2} D. Burgarth, S. Bose, Phys. Rev. A 71, 052315
(2005).
\bibitem{burgarth} V. Giovannetti, D. Burgarth, Phys. Rev. Lett.
96, 030501 (2006).
\bibitem{Avellino} M. Avellino, A.J. Fisher, S. Bose,
Phys. Rev. A 74, 012321 (2006).
\bibitem{bayat} A. Bayat, V. Karimipour, Phys.Rev. A 71, 042330
(2005).
\bibitem{burgarth3} D. Burgarth, S. Bose, Physical Review A 73, 062321
(2006).
\bibitem{jian} J.M. Cai, Z.W. Zhou, G.C. Guo, Phys. Rev. A. 74, 022328 (2006).
\bibitem{osborne} T.J. Osborne, N. Linden Phys. Rev. A 69, 052315
(2004).
\bibitem{subrahamanyan} V. Subrahmanyam Phys. Rev. A 69, 034304
(2004).
\bibitem{chiara} G.De Chiara et.al., Phys. Rev. A 71, 042330
(2005).
\bibitem{clone} R. F. Werner, Phys. Rev. A58, 1827(1998).
\bibitem{Keyl} M. Keyl and R. F. Werner, J. Math. Phys. 40, 3283 (1999).
\bibitem{cryp} A. Acin, N. Gisin, V. Scarani, Quant. Inf. Comp. Vol.3 No. 6, 563
(2003).
\bibitem{tele} G. Rigolin, Phys. Rev. A, 71(2005)032303.
\bibitem{tele2} X. H. Ge and Y. G. Shen, Phys. Lett. B606
(2005)184.
\bibitem{kai} O. Romero-Isart, Kai Eckert,  and A. Sanpera, quant-ph/0610210.
\bibitem{aubert} S. Aubert, C.S. Lam, J.Math.Phys. 45 (2004)
3019-3039.
\bibitem{gisin} N. Gisin, G. Ribordy, W. Tittel and H. Zbinden, Rev.
Mod. Phys. 74, 145 (2002);
\bibitem{ursin} R. Ursin et.al., quant-ph/0607182.
\bibitem{jennewein} T. Jennewein, C. Simon, G.W eihs, H. Weinfurter, A. Zeilinger,
 Phys. Rev. Lett. \textbf{84}, 4729 (2000),quant-ph/9912117.
\bibitem{aspelmeyer}M. Aspelmeyer, T. Jennewein, M. Pfennigbauer, W. Leeb, A. Zeilinger,
IEEE Journal of Selected Topics in Quantum Electronics 1541- 1551,
quant-ph/0305105.
\bibitem{vienna1} K.J.Resch , \emph{et.al} , Opt. Express 13, 202-209
(2005), quant-ph/ 0501008 .
\bibitem{vienna2} A. Poppe, \emph{et.al}
,Opt. Express 12, 3865-3871 (2004) ,quant-ph/0404115.
\bibitem{vidal} G. Vidal, R.F. Werner, Phys. Rev. A 65, 032314
(2002).
\bibitem{plenioneg} M. B. Plenio, Phys. Rev. Lett. 95, 090503
(2005).
\end{thebibliography}
\end{document}